\documentclass[reqno,dvips,12pt]{article}

\usepackage{epsfig,graphicx,amsmath,amssymb,amsfonts}
\newtheorem{theorem}{\qquad Theorem}

\newcommand{\cA}{{\cal A}}
\newcommand{\cM}{{\cal M}}

\newcommand{\cN}{{\cal N}}

\newcommand{\nn}{\nonumber}

\begin{document}

\title{Quasithermodynamics and a Correction to the Stefan--Boltzmann Law}
\author{\textbf{V.~P.~Maslov}}
\date{}

\maketitle

\begin{abstract}
We provide a correction to the Stefan--Boltzmann law and discuss
the problem of a phase transition from the superfluid state into
the normal state.

\end{abstract}

In his book \cite{Shred}, Schr\"odinger notes that the asymptotic
form at a large number of particles $N$ can be obtained with a
precision not better than $\sqrt{N}$. The leading term of the
asymptotic form as $N \to\infty$ with the volume $V \to\infty$ but
with $N/V \to const$ is called the thermodynamic limit. In the
example of the Stefan-Boltzmann radiation law, we show that the
next asymptotic term has the order $N^{2/3}$ and that it is
impossible to achieve a better precision. We also give this term
explicitly. We say that this term, i.e., the limit of the
difference between the exact answer and the thermodynamic limit,
divided by $N^{2/3}$, is the quasithermodynamic limit.

In this note, we present some general considerations about the
thermodynamical limit in statistical physics. First we dwell on
the study of the notion of black body.

The Rayleigh--Jeans formula describing black bodies in classical
physics, valid for low frequencies, was extended in 1900 to high
frequencies in the form of the famous formula due to Planck, who
proposed to consider discrete energies and introduced the
constant that now bears his name. That formula implies, in
particular, the Stefan--Boltzmann formula, which had been
discovered earlier.

The derivation of the Rayleigh--Jeans formula is based on the
Maxwell equation and the Gibbs distribution in classical
mechanics. It also makes use of the complete isotropy of black
emission.

On the basis of the assumption that black body emission is
completely isotropic (see \cite{Landau}, p.205]), we rigorously
obtain a correction to the Planck formula and to the
Stefan--Boltzmann law. In many precise experiments about black
radiation, the necessity of this correction was not noticed. On
the other hand, the exact Maxwell equation for free photons,
which is used to derive the Rayleigh--Jeans law, does not contain
the parameter $h$, and the appearance of a correction to a law
containing this parameter, leads to the problem of carrying over
the correction containing the parameter to the Maxwell equation.

As to the Stefan--Boltzmann law, which reads

\begin{equation}
\label{eq1:u069}
 F =-\frac{4\delta}{3c}\mspace{1mu}V T^4,
\end{equation}
where $F$ is the free energy, $V$ is the volume, $c$ the speed of
light, and $\delta$ the Stefan--Boltzmann constant,
$\delta=\frac{\pi^2k^4}{60h^4c^2}$, it is a fact that here small
discrepancies between theory and experiment were observed. Thus,
the Physical Encyclopedic Dictionary of 1966 \cite{PhED}, p.82
says: ``The experimental value of $\delta _{exp}$ is somewhat
larger than its theoretical value $\delta _{theor}$ obtained by
integrating the Planck formula over the wavelength $\lambda$ (or
the frequency $\nu$). The reasons for this discrepancy are not
quite clear."

The quasithermodynamic correction to this law is given by
\begin{equation}\label{popravka1}
F=-\frac{4\delta}{3c}T^4V - 2\zeta(3) \frac{k^3T^3}{(hc)^2}
V^{2/3},
\end{equation}
where $\zeta$ is the Riemann zeta-function. This result follows
from Theorem 3 in \cite{RJMP_14}, pertaining to number theory.
This general theorem is applicable to objects in economics,
linguistics, and semiotics. In our case, it determines the domain
where most of the choices are positioned that satisfy the
inequality
\begin{equation}\label{2b}
\sum_{i,j,k} (i+j+k)N_{i,j,k} \leq E,
\end{equation}
where $E\gg 1$ is fixed, and the $N_{i,j,k}$  are any natural
numbers (the number of choices corresponding to (\ref{2b}) is
finite). The theorem is roughly formulated as follows (see
\cite{MatZam_80} for the exact formulation): the probability that
the number of choices --
\begin{equation}\label{General_Theor}
 |\sum_{i+j+k \leq s} (i+j+k) N_{i,j,k} - \sum_{i+j+k \leq s}
 \frac{i+j+k}{e^{(i+j+k)b}-1}|\geq \overline {E} N^{\frac12+\varepsilon},
\end{equation}
where $s$ is fixed, is exponentially small as $E \to\infty$; in
other words, there are not too many choices outside the specified
interval. The parameter $b$ is here determined from the equation
\begin{equation}\label{1b}
\sum \frac{i+j+k}{e^{(i+j+k)b}-1} =E,
\end{equation}
and $N$ is the mean value of the sum $
\sum_{i,j,k} N_{i,j,k}$
under condition (\ref{2b}), $\overline {E}=E/N$.

This theorem can be applied to quantum statistics if the
following assumptions are made:

1.  All choices of particle distribution over energy levels are
equiprobable (something of the sort of the equidistribution law,
which we call the "no-preference law").

2.  Blackbody radiation is totally isotropic.

Suppose we have the condition

\begin{equation}
\label{eq2} \sum_{i,j,k}(i+j+k)N_{i,j,k}\le E,
\end{equation}
where $E$ is a constant, $N_{i,j,k}$ are ``equally chosen"
integers, i.e., are equiprobable or equally distributed. Then it
follows from the analog of the theorem in \cite{RJMP_14} that most
of the variants will cluster near the following dependance of the
``cumulative probability":
$$
B_l=\sum_{i+j+k\le l}^lN_{i,j,k},
$$
with $l\leq s, s=\max(1+j+k)$ in inequality (\ref{eq2})

\begin{equation}
 B_l=\sum_{i=1}^l\frac12
\cdot\frac{(i+1)(i+2)}{e^{b i}-1}\mspace{1mu}, \label{eq3:u069}
\end{equation}
where
$b$
is determined from the conditions

\begin{equation}
\label{eq4} \sum_{i=1}^s\frac12\cdot \frac{i(i+1)(i+2)}{e^{b
i}-1}=E,
\end{equation}
if $E\to\infty$.
Denote by $N$ the mean value of the sum
$\sum_{i,j,k}N_{i,j,k}$ under condition (\ref{eq2}).

It turns out that, in this case, most of the variants will also
cluster around the following dependance of ``local energy":
$E_l=\sum_{i+j+k\le l}(i+j+k)N_{i,j,k}$, $E_l\le E$, near
\begin{equation}
\label{eq5:u069} \sum_{i=1}^l\frac12 \cdot\frac{i(i+1)(i+2)}{e^{b
i}-1}\mspace{1mu},
\end{equation}
where $b$ is determined from conditions (\ref{eq4}) if
$E\to\infty$.

The theorem from \cite{MatZam_80} can then be stated similarly to
the theorem from \cite{RJMP_14}.

\begin{theorem}
\label{t1:u069}
 Suppose that all choices of the families
$\{N_i\}$
such that

\begin{equation}
\label{eq6} \sum_{i,j,k}(i+j+k)N_{i,j,k}\le E,
\end{equation}
are equiprobable.
 Then the number of variants
$\cal N$ of families $\{N_{i,j,k}\}$ satisfying (\ref{eq2}) and
(\ref{eq6}) as well as the following additional condition:

\begin{equation}
\label{eq7} \biggl|\sum_{i+j+k\le l}(i+j+k)N_{i,j,k}
-\sum^l_{i=1}\frac12\cdot \frac{i(i+1)(i+2)}{e^{bi}-1}\biggr|
\ge\overline{E}N^{1/2}(\ln N)^{1/2+\varepsilon},
\end{equation}
is less than $(c_1\cal N)/N^m$, where $c_1$ and $m$ are arbitrary
numbers); here $\overline{E}=E/N\sim1/b$.
\end{theorem}

 Notation:
$\cal M$ is the set of all families $\{N_{i,j,k}\}$ satisfying
condition (\ref{eq2}), $\cal N\{\cal M\}$ is the number of
elements of $\cal M$,\enskip $\cal A$ is a subset of $\cal M$
satisfying the condition
$$
\biggl| E_l-\sum_{i=0}^l\frac12
\cdot\frac{i(i+1)(i+2)}{e^{bi}-1}\biggr| \le\Delta,\qquad
l=0,1,\dots,s,
$$
where
$\Delta$ and $b$
are some real numbers not depending on~$l$.

 Denote
$$
\biggl| E_l-\sum_{i=1}^l\frac12
\cdot\frac{i(i+1)(i+2)}{e^{bi}-1}\biggr|=S_l.
$$
Let us recall the scheme of the proof, similar to that of Theorem
3 from \cite{RJMP_14}, in our case.

It is obvious that if $\cal M$ is the number of families
$\{N_{i,j,k}\}$, then
\begin{eqnarray}\label{eq8}
&&\cN\{\cM \setminus \cA\} =  \nn \\
&&\sum_{\{N_{i,j,k}\}} \Bigl( \Theta\bigl\{E-\sum_{i,j,k}
(i+j+k)N_{i,j,k}\bigr\} \delta_{(\sum_{i,j,k}
(i+j+k)N_{i,j,k}),N} \prod^s_{l=0}\Theta\Bigl\{|S_{l}-\Delta|
\Bigr\} \Bigr).
\end{eqnarray}
Here the sum is taken over all integers $N_{i,j,k}$,
$\Theta(\lambda)$ is the Heaviside function, and
$\delta_{k_1,k_2}$ is the Kronecker delta.

 Using the integral representations

\begin{align}
\delta_{NN'} &=\frac{1}{2\pi}\int_{-\pi}^\pi
d\varphi\,e^{-iN\varphi}e^{i N'\varphi}, \label{eq9}
\\
\Theta(y) &=\frac1{2\pi i}\int_{-\infty}^\infty
d\lambda\,\frac1{\lambda-i}\mspace{1mu} e^{by(1+i\lambda)}.
\label{eq10}
\end{align}

and performing the standard regularization, we obtain
\begin{equation}
\int_{0}^\infty
dE\,\Theta\biggl(E-\sum_{i,j,k}(i+j+k)N_{i,j,k}\biggr)e^{-b E}
=\frac{e^{-b\sum_{i,j,k}(i+j+k)N_{i,j,k}}}{b}\mspace{1mu}.
\label{eq11:u069}
\end{equation}

Denote
$$
Z(b,N)=\sum_{\{N_{i,j,k}\}}
e^{-b\sum_{i,j,k}(i+j+k)N_{i,j,k}},
$$
where the sum is taken over all
$N_{i,j,k}$.
 Further, introduce the notation
$$
\zeta_l(i\alpha,b)
=\prod_{i=1}^l\xi_i(i\alpha,b),\qquad
\xi_j(i\alpha,b)
=\frac1{(1-e^{i\alpha-bj})^{j(j+1)/2}}\mspace{1mu},\quad
j=1,\dots,l.
$$
It follows from (\ref{eq9}) that

\begin{equation}
 Z(b,N)=\frac{1}{2\pi}\int_{-\pi}^\pi
d\alpha\,e^{-iN\alpha}\zeta_s(i\alpha,b); \label{eq12}
\end{equation}
hence

\begin{equation}
\begin{split}
\label{eq13:u069} \cal N\{\cal M\setminus\cal A\}
&\le\biggl|\frac{e^{b E}}{i(2\pi)^2} \int_{-\pi}^\pi\biggl[
\exp(-iN\varphi)
\\&\qquad\qquad\qquad \times
\sum_{\{N_{i,j,k}\}}\biggl(\exp\biggl\{
\biggl(-b\sum_{i,j,k}(i+j+k)N_{i,j,k}\biggl)
+(i\varphi)N_{i,j,k}\biggr\}\biggr)\biggr]\,
d\varphi
\\&\qquad\qquad \times
\prod_{l=0}^s\Theta(|S_{l}-\Delta|)\biggr|,
\end{split}
\end{equation}
where
$b$
is a real parameter for which the series converges.

Estimating the right-hand side, carrying the absolute value sign
under the integral and then further under the sum, we obtain
after integration over~$\varphi$
\begin{eqnarray}\label{eq14}
&&\cN\{\cM \setminus \cA\} \leq \frac{e^{b E }}{2\pi}
\sum_{\{N_{i,j,k}\}}\exp\{-b\sum_{i,j,k} (i+j+k)N_{i,j,k} \}\times \nn \\
&& \times\prod_{l=0}^s\Theta (|S_{l}-\Delta|).
\end{eqnarray}

 From the inequality for the hyperbolic cosine
$\operatorname{cosh}(x)=(e^x+e^{-x})/2$
\begin{equation}
2^s\prod_{l=0}^s\operatorname{cosh}(x_l) \ge e^\delta\qquad
\forall\,x_l{:}\quad \sum^s_{l=0}|x_l|\ge\delta\ge0, \label{eq15}
\end{equation}
it follows that, for all positive~$c$ and $\Delta$, we can write
(compare ~\cite{10:u069},~\cite{11:u069})

\begin{equation}
\prod_{l=0}^s\Theta(|S_{l}-\Delta|)
\le2^se^{-c\Delta}\prod_{l=0}^s
\operatorname{cosh}\biggl(c\sum_{i+j+k\le l}(i+j+k)N_{i,j,k}
-c\psi_{b}\biggr), \label{eq16}
\end{equation}
where

$$
\psi_b=\sum_{i=1}^l\frac12
\cdot\frac{i(i+1)(i+2)}{e^{bi}-1}\mspace{1mu}.
$$

 Thus, we obtain

\begin{equation}
\begin{split}
\cal N\{\cal M\setminus\cal A\} &\le e^{-c\Delta}\exp(bE)
\sum_{\{N_{i,j,k}\}}\exp\biggl\{
-b\sum_{i,j,k}(i+j+k)N_{i,j,k}\biggr\}
\\&\qquad\qquad \times
\prod_{l=0}^s\operatorname{cosh}\biggl(\mspace{1mu}
\sum_{i+j+k\le l}c(i+j+k)N_{i,j,k}
-c\psi_{b}\biggr)
\\&
=e^{b E}e^{-c\Delta} \prod_{l=0}^s\bigl\{
\zeta_{l}(0,b-c)\exp(-c\psi_{b})
+\zeta_{l}(0,b+c)\exp(c\psi_{b})\bigr\}. \label{eq17}
\end{split}
\end{equation}
Let us apply Taylor's formula to
$\zeta_l(0,b\pm c)$.
 There exists a
$\gamma<1$
such that
$$
\ln(\zeta_l(0,b\pm c))=\ln\zeta_l(0,b)\pm
c(\ln\zeta_l)'_b(0,b)+\frac{c^2}2(\ln\zeta_l)^{''}_b
(0,b\pm\gamma c).
$$
 Obviously,
$$
\frac{\partial}{\partial b}\ln\zeta_l\equiv-\psi_b.
$$
 Let us put
$c=\Delta/D(0,b)$,
where
$D(0,b)=(\ln\zeta_l)_b''(0,b)$
is positive for all
$b$
and monotonically decreases as
$b$
increases.
 The right-hand side of (\ref{eq17}) does not exceed

$$
2^\gamma e^{bE}\prod_{p=0}^s\zeta_{l_p}(0,b)
e^{-(\Delta_l^2)/|D(0,b)|}
+\frac{\Delta_l^2D(0,b-\gamma\Delta/D(0,b))}
{2(D(0,b))^2}\mspace{1mu}.
$$

 As in \cite{RJMP_14}, we obtain
\begin{equation}\label{eq18}
\cN(\cM\setminus \cA)\leq   e^{b E}
\zeta_l(0,b)e^{-{\varepsilon\Delta^2}/{D(0,b)}}.
\end{equation}

Therefore, in the interval from $-\pi$ to $\pi$ over which the
integral (\ref{eq12}) is taken, the only contribution comes from a
neighborhood of the point $\alpha =\nobreak 0$.
 Let us compute the integral (\ref{eq12}) by the Laplace method with precision up to
$N^{-m}$ and then apply all the subsequent arguments from the
proof of Theorem 3 in \cite{RJMP_14}.

In order to estimate $\zeta_s(0,\beta)$ from below, we can use
the exact asymptotics obtained by Krutkov \cite{Krutkov} for the
three-dimensional oscillator. In the case of a general spectrum
$\lambda_n$, the formula obtained by the saddle-point method
meets with considerable difficulties due to the fact that an
infinite number of saddle points can appear in certain concrete
examples (Koval', private communication).

Note that the sum over all variants satisfying inequality
(\ref{eq2}) may be interpreted as a discrete continual ``path
integral", the paths being the variants. The asymptotic leading
term of the continual integral is concentrated, as a rule, near
one principal ``trajectory" (this is the Laplace method for the
continual integral), and rapidly tends to zero outside a
neighborhood of this trajectory. This is expressed by the
statement of the theorem. The size of this neighborhood (in our
case $\overline{E}O(N^{1/2}(\ln N)^{1/2+\varepsilon})$)
determines the limiting precision with which it makes sense to
compute this ``principal trajectory".


In this connection, we propose the following somewhat modified
conjecture about the Schr\"odinger rule that he qualified as a
``law of nature".  If a certain number of particles, molecules,
genes in a chromosome is equal to $N$, then one can obtain a
statistical law with precision of no more than $O(\sqrt{N\ln N})$.
The mathematical meaning of this conjecture is that estimate
(\ref{eq7}) cannot be improved by more than $\varepsilon$.

Since $E\to \infty$ in (\ref{eq4}), it follows that we can put
$s=\infty$ in the sum

$$
\sum_{i=1}^\infty\frac{i(i+1)(i+2)}{2(e^{bi}-1)}\mspace{1mu},
$$
we can apply the Euler formula of the form
\begin{equation}
\begin{gathered}
\begin{aligned}
\sum_{b>n>a}f(n)
&=\int_a^bf(x)\,dx+\rho(b)f(b)-\rho(a)f(a)
\\&\qquad
+\sigma(a)f'(a)-\sigma(b)f'(b)
+\int_a^b\sigma(x)f''(x)\,dx
\end{aligned}
\\
\rho(x)=\frac12-\{x\},\qquad
\sigma(x)=\int_0^x\rho(t)\,dt.
\end{gathered}
\label{eq19:u069}
\end{equation}
 Since
$\sigma(x)\le1/8$,
it is easy to see that
$$
E=\frac{\zeta(4)}{12}\mspace{1mu}b^{-4}
+\frac34\mspace{1mu}\zeta(3)b^{-3}+O(b^{-2}),
$$
where $\zeta(x)$ is the Riemann zeta function. Therefore, $b\to
\infty$. As is known, $\zeta(4)=\pi^4/90$.

Similarly, in the right-hand side of (\ref{eq7}), we can pass to
integrals for sufficiently large values of $l$.

In view of (\ref{eq7}), the term $O(b^3)$ is greater than
$\overline{E}\sqrt{N\ln N}$ and its calculation makes no sense.

Similar estimates for $N$, as can be seen from Theorem 1 from
\cite{MatZam_80}, also make possible the computation of the
asymptotic (in $b$) terms $O(b^{-3})$ and $O(b^{-2})$ only. The
leading (first) term of the asymptotics will be called the {\it
thermodynamical limit} and the second one, the {\it
quasithermodynamical limit}.

Now let us consider a system of $N$ three-dimensional
noninteracting oscillators of the same frequency $\omega_0$:

\begin{equation}
\label{eq20:u069}
-\frac{h^2}{2m}\Delta\Psi_n(x)
-\omega_0^2|x|^2\Psi_n(x)
=n\Psi_n(x),\qquad
x\in\mathbb R^3.
\end{equation}

In order to obtain the leading term of the Stefan--Boltzmann law
for this system of oscillators, we must set

$$
b=\frac{\omega_0\hbar}{kT}\mspace{1mu};
$$
here
$\hbar$
is the Planck constant,
$k$
is the Boltzmann constant,
$T$
the temperature,
$\omega_0$
the frequency, which equals
$\omega_0=c/\sqrt[3]V$,
where
$c$
is the speed of light, and
$V$
is the volume.
 We assume that the oscillators are completely isotropic, so that the frequency
$\omega_0$
is the same in all directions.
 Oscillations of frequency
$\omega$
greater than
$\omega_0$
do not exist at the given temperature, while the frequencies
$\omega < \omega_0$
give a considerably lesser number of variants and may be neglected.

 As a result, for the correction to the Stefan--Boltzmann law, we obtain a quasithermodynamical term of the form

\begin{equation}
\label{eq21:u069}
 F=-\frac{4\delta}{3c}\mspace{2mu}T^4V
-\frac{12\hbar\delta}{k}
\cdot\frac{\zeta(3)}{\zeta(4)}\mspace{2mu}T^3V^{2/3},
\end{equation}
where
$F$
is the free energy,
$\delta=\pi^2k^4/(60\hbar^3c^2)$
is the Stefan--Boltzmann constant,
$V$
is the volume,
$c$,
the speed of light,
$\hbar
$ is the Planck constant,
$k$
is the Boltzmann constant, and
$T$
is the temperature.

It has {\it always} been observed that the experimental values of
the Stefan--Boltzmann constant are larger than their theoretical
values. Now the reason for this discrepancy is clear: the
correction term specified above explains this discrepancy.

However, the main effect of quasithermodynamics occurs when there
is no thermodynamocal phase transfer, but there is a
quasithermodynamical one. It is precisely such an effect that was
obtained by the author in the study of a Fermi-gas without the
additional assumption on the existence of Cooper pairs. It was
assumed that the number of particles $N$ tends to $\infty$, the
particle interaction is pairwise, as in Helium 4, i.e, repulsive
at short distances and attractive at large ones. At the same
time, the potential $V(r_i - r_j)$ is mainly a short distance one.
The corresponding spectrum has the form

\begin{equation}
\label{eq22}
 E_l=\frac{\hbar^2l^2}{2m}
+\widetilde{V}(l)-\widetilde{V}(0),
\end{equation}
where
$\widetilde V(l)$
is the Fourrier transform of
$V(z)$.

As $N$ tends to infinity and the volume $V$ tends to infinity in
such a way that $N/V\to\operatorname{const}$, the potential tends
to the $\delta$-function and therefore tends to the spectrum of
an ideal gas. Superfluidity arises only in the
quasithermodynamical limit. In this limit, as a rule, the Landau
curve arises. Thus, the phase transfer to the superfluid state
occurs in the quasithermodynamics of this ``model". Note that
this model does not contain any additional physical interactions:
it is a ``model without a model", since the antisymmetric
solution of the $N$-particle Schr\"odinger equation with the
ordinary pairwise interaction cannot be regarded as a ``model".
This is an ordinary mathematical problem.

Let us consider a Bose-gas. Bogolyubov proposed the following
spectrum in this case:
\begin{equation}
\label{eq23}
 E_l=\sqrt{\biggl(\frac{\hbar^2l^2}{2m}
+\widetilde{V}(l)\biggr)^2-\widetilde{V}(l)^2}\,,
\end{equation}
which also yields the Landau curve in quasithermodynamics. But
unlike (\ref{eq22}), in the thermodynamical limit it gives
$$
\lim_{N/V\to\mathrm{const},\,N\to\infty}E_l
=\sqrt{\biggl(\frac{\hbar^2l^2}{2m}
+\widetilde V(0)\biggr)^2-\widetilde V(0)^2}\,,
$$
and superfluidity is preserved (the critical Landau speed in the
thermodynamical limit is not zero, as it is in the case of the
spectrum (\ref{eq22})). However, the photon part of the spectrum
disappears in the thermodynamical limit and appears only in
quasithermodynamics.

\textbf{Remark.} Bogolyubov's work may be rigorously founded under
certain additional conditions, provided one considers the
problem, as Bogolyubov did, on the three-dimensional torus.
However, the passage to the limit from the torus to
three-dimensional space is erroneous: an everywhere dense point
spectrum appears in the limit.

The author obtained a modification of formula (\ref{eq23}) for the
case in which the liquid flows through a capillary of radius $r$.
The spectrum in that case has the form

\begin{equation}
\label{eq24:u069}
\begin{split}
 E_l&=\Biggl[
\frac{1}{2}\biggl(a(l^2-k_2^2)
+\frac{\widetilde{V}(l-k_2)-\widetilde{V}(k_2)}{2}
\biggr)^2
+\frac{1}{2}\biggl(a(l_1^2-k_2^2)
+\frac{\widetilde{V}(l-k_2)-\widetilde{V}(k_2)}{2}
\biggr)^2
\\&\qquad
+\biggl(\frac{\widetilde{V}(l+k_2)+\widetilde{V}(k_2)}{2}
\biggr)^2-\biggl(\frac{\widetilde{V}(l-k_2)+\widetilde{V}(l+k_2)}{2}
\biggr)^2
\\&\qquad
+\frac{1}{2}\bigl(a(l_1^2+l^2-2k_2^2)+\widetilde{V}(l-k_2)
-\widetilde{V}(k_2)\bigr)
\\&\qquad\qquad \times
\sqrt{a^2(l_1^2-l^2)^2
+2(\widetilde{V}(l+k_2)+\widetilde{V}(k_2))^2}\,
\Biggr]^{1/2},
\end{split}
\end{equation}
where
$$
a=\frac{h^2}{2m}\mspace{1mu},\qquad
l_1=l+2k_2,
$$
while
$k_2=2\pi n/r$,
where
$n=1,2,\dots$\,.

 When
$r=\infty$, and hence $k_2=0$, we obtain the Bogolyubov formula
(\ref{eq23}). In the limit as $r\to0$, $k_2\to\infty$,
$l=(l_0,-k_2)$ ($l_0$ is the component along the direction of
flow), we obtain

$$
\lim_{r\to0}E_{l_0} =\sqrt{a^2l_0^4+a|\widetilde V(l_0)|l_0^2}\,.
$$

These results, just as the Bogolyubov formula, are valid in the
thermodynamical limit with the quasithermodynamical correction.

 \section*{{Acknowledgments}}

The author would like to express deep gratitude to
A.~A.~Karatsuba.

\end{document}